\documentclass[a4paper,12pt]{article}
\usepackage{amssymb}
\usepackage{graphicx}
\usepackage{booktabs}
\usepackage{array}
\usepackage{multirow}
\usepackage{hyperref}
\usepackage{cite}
\usepackage{appendix}
\usepackage{amsmath}
\usepackage{geometry}
\usepackage{color}
\usepackage{xcolor}
\usepackage{titlesec}
\usepackage{soul}
\soulregister\cite7
\usepackage{enumitem}
\usepackage{float}
\usepackage{caption}
\usepackage{tabularx}
\usepackage{makecell}

\begin{document}
\title{\bf{An overview of scale invariance in proton structure with holographic insights}}

\author{ $ \mathrm{Akbari \; Jahan}^{\star} $ \\ Department of Physics \\  North Eastern Regional Institute of Science and Technology,\\ Nirjuli-791109, Arunachal Pradesh, India.\\ $ {}^{\star} $Email: akbari.jahan@gmail.com}

\date{}

\maketitle

\begin{abstract}
The concept of self-similarity in the internal structure of the proton, rooted in scale invariance and fractal geometry, provides an intriguing framework for understanding the behaviour of parton distribution functions (PDFs), particularly in the small \textit{x} region probed in deep inelastic scattering (DIS). Phenomenological models based on self-similarity have been shown to reproduce key features of experimental data, suggesting that recursive scaling patterns may play an important role in partonic dynamics. In this work, we present an overview of scale-invariant descriptions of proton structure, focusing on self-similar models developed in earlier studies and their phenomenological implications for structure functions and parton distributions. We then explore possible conceptual connections between these fractal-inspired descriptions and modern holographic approaches to QCD, particularly within the framework of light-front holographic QCD. By comparing the scaling behaviour appearing in phenomenological models with the geometric structure underlying holographic QCD, we highlight qualitative correspondences that suggest a broader role of scale invariance in proton structure. Although the connection remains interpretive rather than derivational, it offers a complementary perspective of how fractal-like scaling observed in DIS may relate to geometric scaling in holographic descriptions of QCD.\\

\textbf{Keywords \,:} Scale invariance, Self-similarity, Parton distribution functions, Holographic QCD.\\


\end{abstract}

\section{Introduction}
The proton, though one of the most familiar particles in the Standard Model, remains an object of intense study. Understanding the internal dynamics of the proton remains a central challenge in hard scattering processes. Its internal structure is described in terms of quarks and gluons bound together by the strong interaction. Quantum Chromodynamics (QCD), the gauge theory of strong interactions, provides the fundamental framework, but it becomes computationally intractable in the non-perturbative regime. Parton distribution functions (PDFs), transverse momentum dependent distributions (TMDs), and structure functions such as $F_2(x,Q^2)$ and $F_L(x,Q^2)$ offer empirical windows into partonic dynamics. PDFs encode the probability of finding a parton carrying momentum fraction \textit{x} at resolution scale $Q^2$. They are indispensable inputs for calculations at hadron colliders such as the Large Hadron Collider (LHC). In proton-proton collisions, PDFs provide the nonperturbative input required to calculate hard-scattering cross sections. Modern event generators such as PYTHIA combine PDFs with parton showers, multiparton interactions, colour reconnection models, and hadronization mechanisms to describe the complete final state. Consequently, improving our understanding of proton structure remains important not only for DIS phenomenology but also for precision modelling of collider observables at the LHC \cite{Alrebdi}. Over the past decades, experimental advances in deep inelastic scattering (DIS) have unveiled complex patterns in the distributions of quarks and gluons, hinting at an underlying self-similar or fractal-like organization. Such self-similarity suggests that the proton’s structure maintains invariant properties under scale transformations, a feature reminiscent of fractal geometries encountered in diverse physical systems. Among the most profound areas where self-similarity plays a central role is in modeling proton structure and high-energy interactions, including those at the LHC. In Ref. \cite{Lastovicka, Gogoi}, self-similarity based models were introduced to describe proton structure, leveraging concepts from fractal geometry. Our earlier works \cite{AJ1, AJ2, AJ3, AJ4, AJ5, AJ6} delve into self-similarity based models to describe the proton’s internal structure, the distribution of partons (quarks and gluons), and the associated longitudinal structure function $F_L(x,Q^2)$  at small \textit{x} \cite{Blumlein}. $F_L(x,Q^2)$  is a key quantity in DIS and is sensitive to both quark and gluon distributions as given in Eq.(\ref{eqn:fl}) \cite{Altarelli}:

\begin{equation}
\label{eqn:fl}
F_{L}\left(x,Q^{2}\right) = \frac{\alpha_s(Q^2)}{4 \pi} \left[\frac{4}{3} \int_x^1 \frac{dy}{y} \left(\frac{x}{y} \right)^2   F_{2}\left(y,Q^{2}\right) + 2 \sum_q e_q^2  \int_x^1 \frac{dy}{y} \left(\frac{x}{y} \right)^2  \left(1- \frac{x}{y}  \right) g\left(y,Q^{2}\right)  \right]
\end{equation}

In perturbative QCD, the longitudinal structure function $F_{L}\left(x,Q^{2}\right)$ receives contributions beyond the naive quark-parton model and is particularly sensitive to the gluon content of the proton at small \textit{x}. At leading order in $\alpha_s$, $F_L$ may be expressed through the Altarelli-Martinelli relation \cite{Altarelli, Rezaei}, which involves convolutions of coefficient functions with both the quark and gluon parton distributions. In Eq.(\ref{eqn:fl}), $g\left(y,Q^{2}\right) $ denotes the gluon parton distribution function evaluated at momentum fraction \textit{y} and resolution scale $Q^2$, while $e_q$ represents the electric charge of quark flavour \textit{q}. The variable \textit{y} is the integration variable associated with the momentum fraction carried by the parent parton entering the hard scattering process. In Ref.\cite{AJ1, AJ6}, $F_L$ in the context of self-similar and fractal-inspired proton structure models are discussed. The concept of self-similarity in proton structure was pioneered in models that capture the recursive scaling of parton densities, notably in several literature \cite{Lastovicka, Gogoi, AJ1, AJ2, AJ3, AJ4, AJ5, AJ6, Baishali1, Baishali2, Baishali3, Mohsenabadi}. These models yield novel parameterizations of PDFs and structure functions, particularly effective in the small-\textit{x} domain. Perturbative QCD underlies this behavior, with DGLAP evolution \cite{DGLAP1, DGLAP2, DGLAP3, DGLAP4} governing moderate \textit{x} scaling and BFKL formalism \cite{BFKL1, BFKL2, BFKL3, BFKL4} describing the small \textit{x} regime. HERA experiments at DESY greatly extended the accessible kinematic region, revealing a steep rise of $F_2(x,Q^2)$ at $x \lesssim 10^{-3}$, signaling gluon dominance \cite{H1}. Global PDF analyses such as NNPDF \cite{NNPDF, Ball2} and MMHT \cite{MMHT, Harland2} now combine DIS, Drell-Yan, jet, and vector boson data to produce precision PDFs for collider physics. Beyond these standard parameterizations, alternative models have sought to embed deeper principles, such as scale invariance, into proton structure.\\

Self-similarity connects naturally to the language of holography, where scale transformations in boundary field theories correspond to geometric transformations in a higher-dimensional bulk spacetime \cite{Tehrani}. Simultaneously, advances in holographic QCD and AdS/CFT correspondence have opened new pathways to explore hadronic properties using higher-dimensional gravity duals. This paper explores the convergence of these two frameworks, exploring whether qualitative similarities exist between self-similar descriptions and holographic approaches. Our study also aims to provide a comprehensive account of the theoretical underpinnings of self-similarity in proton structure and its holographic implications. The purpose of this paper is not to introduce new holographic calculations or quantitative predictions, but rather to provide a critical and coherent perspective on how phenomenological self-similar descriptions of proton structure may be interpreted within a holographic setting. By comparing fractal-inspired PDFs with holographic descriptions, we aim to offer a cohesive framework for understanding the proton’s scale-invariant properties and their broader relevance in physics. Throughout this review, a distinction is maintained between three levels of description. First, established results from QCD and deep inelastic scattering experiments provide the empirical and theoretical foundation for our understanding of proton structure. Second, phenomenological self-similar models are discussed as effective parametrizations that reproduce certain observed scaling features of parton distributions. Third, possible connections with holographic QCD are considered at a qualitative and interpretive level. No equivalence between these frameworks is implied, and the discussion of holography is intended primarily to highlight conceptual analogies rather than establish a formal derivation. Section 2 reviews the essential features of self-similar models of the proton and their phenomenological implications. Section 3 discusses holographic descriptions of the proton, with an emphasis on light-front holographic QCD and its treatment of scale invariance. Section 4 provides an interpretive synthesis of these approaches, emphasizing conceptual correspondences rather than derived results. Section 5 summarizes the key findings and highlights the possible extensions of this approach to more comprehensive descriptions of parton dynamics.

\section{Self-similar and scale-invariant descriptions of proton structure}

\begin{figure}[h]
\centering
\includegraphics[scale=0.75]{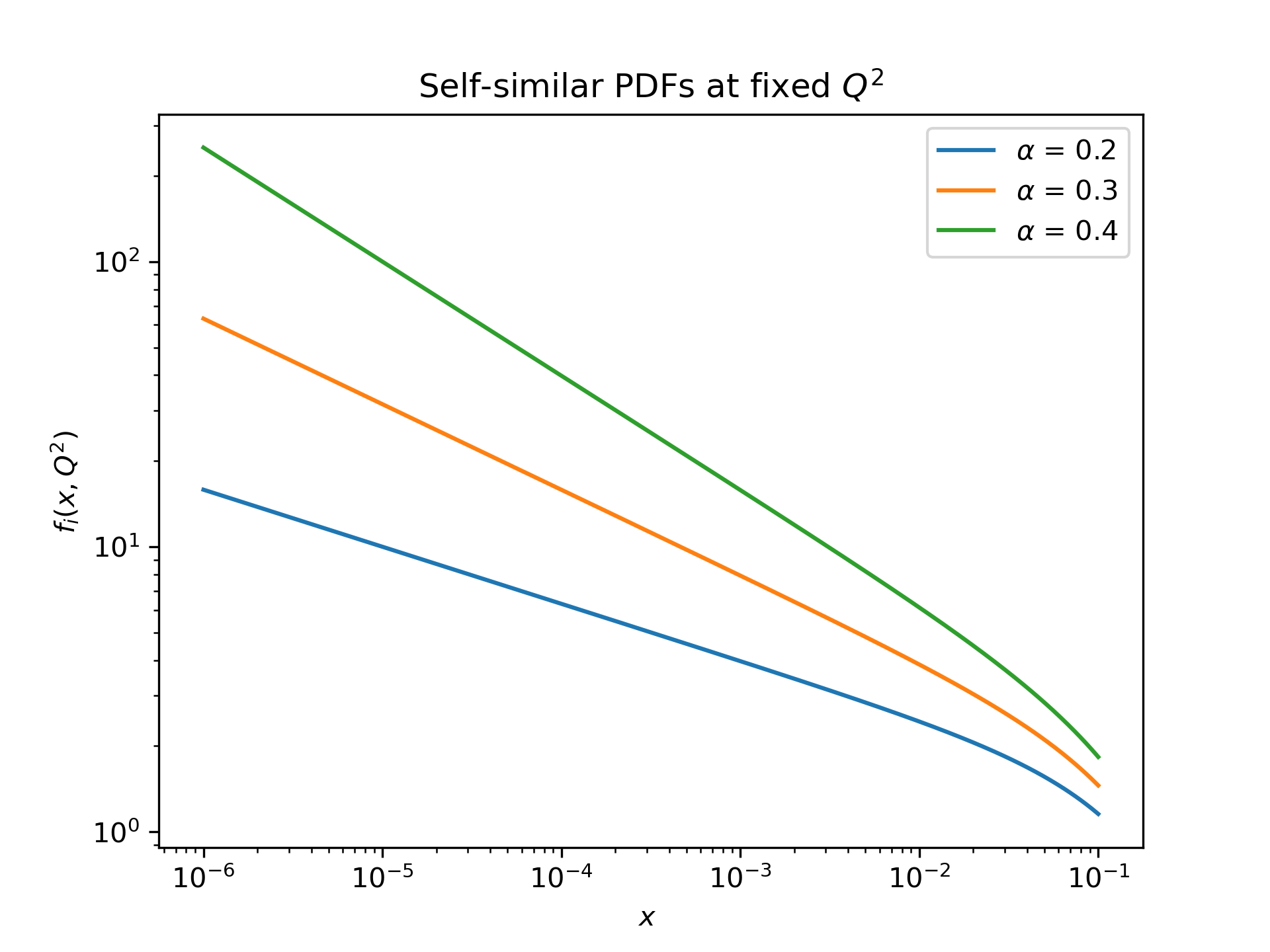}
\caption{Illustrative behaviour of the self-similar parton density $f_i(x,Q^2)$ as a function of \textit{x}, based on the functional form of Eq.(\ref{eqn:fi}). The curves are generated using illustrative parameter values $\alpha = 0.2, 0.3, 0.4$ with $\beta = 3$. No specific value of $Q_0^2$ is used because the figure is intended solely to demonstrate the qualitative scaling behaviour characteristic of self-similar descriptions and does not represent a fit to experimental data.}
\label{Figure1}
\end{figure}

The self-similar descriptions reviewed in this section should be understood as phenomenological models motivated by observed scaling behaviour and constrained by experimental data, rather than as results derived directly from first-principles QCD. Self-similarity, the invariance of a structure under scale changes, implies in hadronic physics that the proton's internal composition exhibits recursive or fractal-like patterns when probed at different momentum scales. The idea that the proton might display such self-similar organization of its partonic constituents has proven to be phenomenologically useful in describing DIS data at small \textit{x}. In this approach, PDFs are represented by scale-free or fractal-inspired forms that reflect an underlying invariance of the proton structure under transformations of \textit{x} and  $(Q^2)$ \cite{Lastovicka, Gogoi, AJ1, AJ2, AJ3, AJ4, AJ5, AJ6, Baishali1, Baishali2, Baishali3}. A convenient expression for such scaling behaviour is

\begin{equation}
\label{eqn:fi}
f_{i}\left(x,Q^{2}\right) = x^{-\alpha} (1-x)^{\beta} \left(\frac{Q^2}{Q_0^2} \right)^{\gamma(x)}
\end{equation}

where $\alpha$ and $\beta$ control the small \textit{x} and large \textit{x} behaviour respectively, and $\gamma(x)$ governs the evolution of the distribution with the resolution scale $Q^2$ (see Figure \ref{Figure1}).

\begin{figure}[h]
\centering
\includegraphics[scale=0.75]{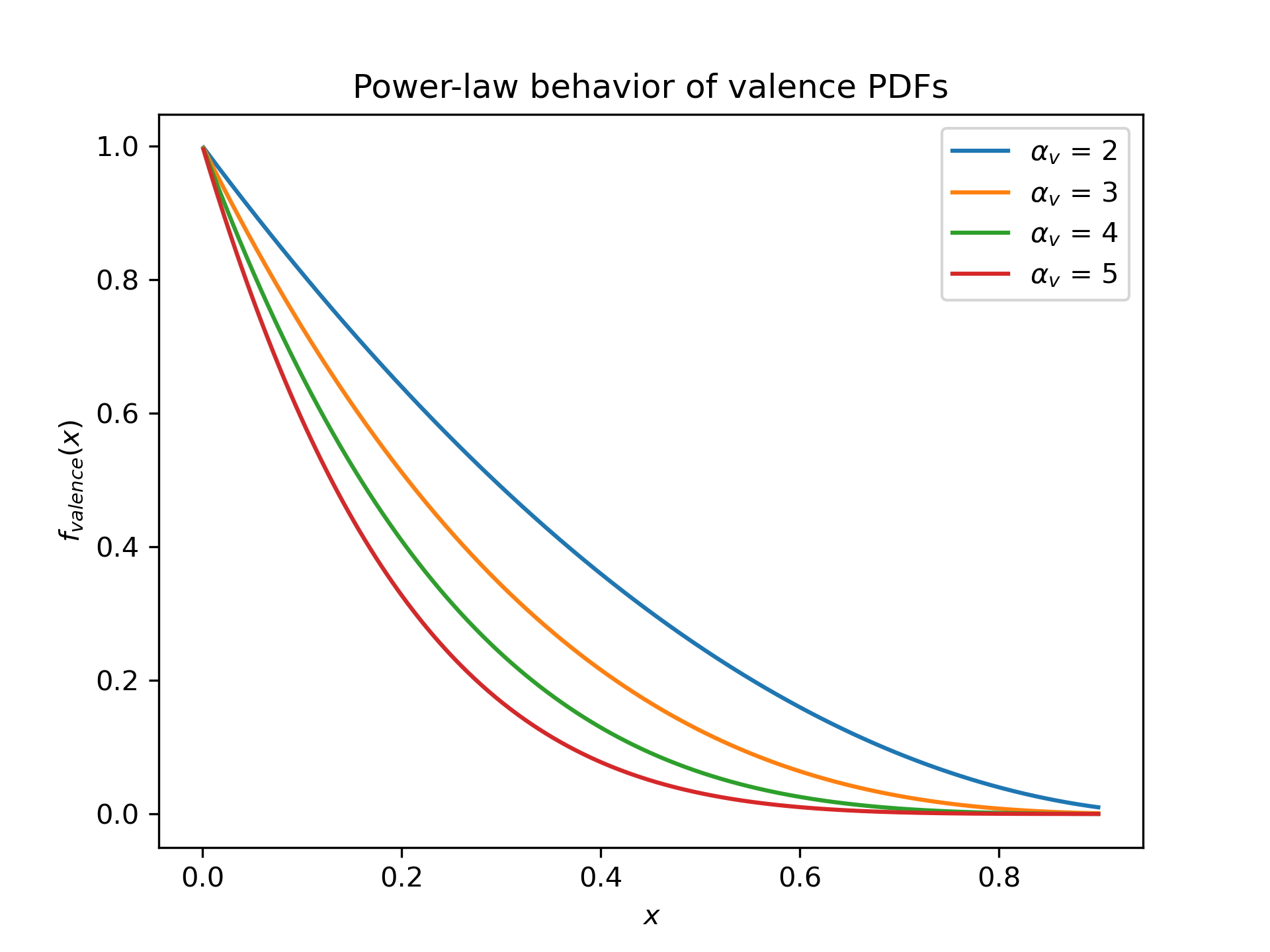}
\caption{Qualitative behaviour associated with the functional form appearing in Eq.(\ref{eqn:fvalence}), showing the suppression of the distribution as $x \rightarrow 1$.}
\label{Figure2}
\end{figure}

The parameterizations discussed in this section are phenomenological in nature and are constrained primarily through fits to experimental data rather than derived from first principles QCD. The power-law dependence in \textit{x} embodies the notion that each partonic scale statistically resembles smaller sub-structures. The $Q^2$-dependence captures how the parton cascade evolves with increasing resolution. The longitudinal structure function $F_L(x,Q^2)$, defined in Eq.(\ref{eqn:fl}), is particularly sensitive to the gluon component at small \textit{x}. The steep increase of $F_L$ and $F_2$ at small \textit{x} is generally interpreted as evidence for a rapidly growing gluon density. At sufficiently small \textit{x}, nonlinear phenomena such as gluon saturation may become relevant, although the onset and quantitative impact of such effects remain active areas of investigation \cite{Albacete}. Self-similar models reproduce this rise naturally because the fractal exponents $\alpha$ and $\gamma(x)$ amplify the small \textit{x} contribution through recursive scaling. At large \textit{x}, where valence quarks carry most of the proton's momentum, the distributions are expected to fall off according to a power law:
 
\begin{equation}
\label{eqn:fvalence}
f_{valence}(x) \sim \left(1-x \right)^{\alpha_v}
\end{equation}
where $\alpha_v$ is a positive constant (see Figure \ref{Figure2}).\\

This behavior arises from the limited phase space when one quark carries nearly all of the momentum. Within a self-similar framework, the same scaling pattern persists across energy scales and can be written phenomenologically as

\begin{equation}
\label{eqn:fq}
f_q \left(x,Q^{2}\right) \sim \left(1-x \right)^{\alpha (Q^2)}  f_q \left(x,Q_0^{2} \right)
\end{equation}

where the exponent $\alpha (Q^2)$ varies smoothly with the scale. The shape of the parton distribution thus remains similar under changes in $Q^2$, expressing continuous self-similarity in the large \textit{x} region. A more formal link with QCD evolution arises from the DGLAP equation \cite{DGLAP1, DGLAP2, DGLAP3, DGLAP4},

\begin{equation}
\label{eqn:DGLAP}
Q^2 \frac{\partial F_2 \left(x,Q^2 \right)}{\partial Q^2} = \int_x^1 \frac{dy}{y} P \left( \frac{x}{y} \right) F_2 \left(y,Q^2 \right)
 \end{equation}
 
 which governs the scale dependence of the structure functions. Eq.(\ref{eqn:DGLAP})describes how structure functions evolve with $Q^2$, expressing the branching of partons across momentum scales. Eq.(\ref{eqn:DGLAP}) is intended as a schematic representation of perturbative QCD evolution. In a complete QCD treatment, small \textit{x} evolution involves coupled singlet-quark and gluon distributions governed by the DGLAP equations \cite{DGLAP1, DGLAP2, DGLAP3, DGLAP4}. This integral equation describes a branching process in which partons at one resolution generate a distribution of offspring at higher resolution, essentially a continuous form of the fractal cascade. This resemblance should be understood as formal and phenomenological; perturbative QCD evolution is not inherently fractal in a strict mathematical sense. From a phenomenological viewpoint, this evolution bears a close resemblance to the recursive processes that motivate fractal-inspired descriptions of PDFs. In this sense, the perturbative evolution of PDFs through DGLAP equation bears a formal resemblance to the recursive structure that motivates self-similar scaling descriptions, although it arises from perturbative QCD dynamics rather than an explicit fractal construction. To maintain consistency with physical constraints, the self-similar PDFs must satisfy the momentum-sum rule,

\begin{equation}
\label{eqn:sumrule}
\int_0^1  x \left[ \sum_q f_q \left(x,Q^2 \right) + f_g  \left(x,Q^2 \right) \right] dx = 1
 \end{equation}

ensuring that all partons collectively carry the proton's total momentum. This condition serves to constrain the gluon distribution $f_g (x,Q^2)$ once the quark densities are fixed by fits to $F_2(x,Q^2)$ and $F_L(x,Q^2)$. Several fractal-inspired parametrizations have embodied these principles. The early works of Ref. \cite{Lastovicka, Gogoi} proposed functional forms that successfully reproduced DIS data at small \textit{x} using a minimal set of scaling parameters. Later extensions \cite{AJ1, AJ2, AJ3, AJ4, AJ5, AJ6, Baishali1, Baishali2, Baishali3} refined the approach to include momentum-fraction analyses of quarks and gluons, self-similar double parton distributions, and consistency with theoretical limits such as the Froissart bound \cite{Froissart}. Taken together, these studies demonstrate that self-similarity provides not only an empirical description but also a conceptual bridge between nonperturbative QCD and the observed scaling laws. The self-similar perspective implies that the proton's internal dynamics possess a recursive symmetry across \textit{x} and $Q^2$, motivating its extension to holographic frameworks where boundary scale transformations correspond to bulk geometry. The next section explores this connection through the lens of light-front holographic QCD.

\section{Holographic proton models}

Holographic QCD provides a geometric realization of scale transformations in strongly interacting systems. QCD is not exactly conformal, but it exhibits approximate scale invariance at high energies. The scaling behavior observed at small \textit{x} motivates a holographic treatment. In this appraoch, hadronic properties can be modeled through a higher-dimensional dual gravity theory in which bulk field modes encode the dynamics of partons and their spatial distributions \cite{Polchinski, Erlich, Erlich2}. Among various formulations, light-front holographic QCD provides a particularly direct connection between the geometry of $\mathrm{AdS_5}$ space and hadronic dynamics \cite{Brodsky, Brodsky2}. It maps the fifth-dimensional coordinate of AdS space to the invariant transverse separation between partons on the light front, establishing a clear correspondence between spatial depth in the bulk and the boundary resolution scale \textit{Q}. The resulting light-front wavefunctions naturally reproduce the power-law scaling of form factors at large $Q^2$ while maintaining confinement at long distances. Light-front holographic QCD has thus emerged as a robust framework for describing hadronic structure functions, generalized parton distributions, and other observables related to the internal composition of the proton.\\

\begin{figure}[h]
\centering
\includegraphics[scale=0.75]{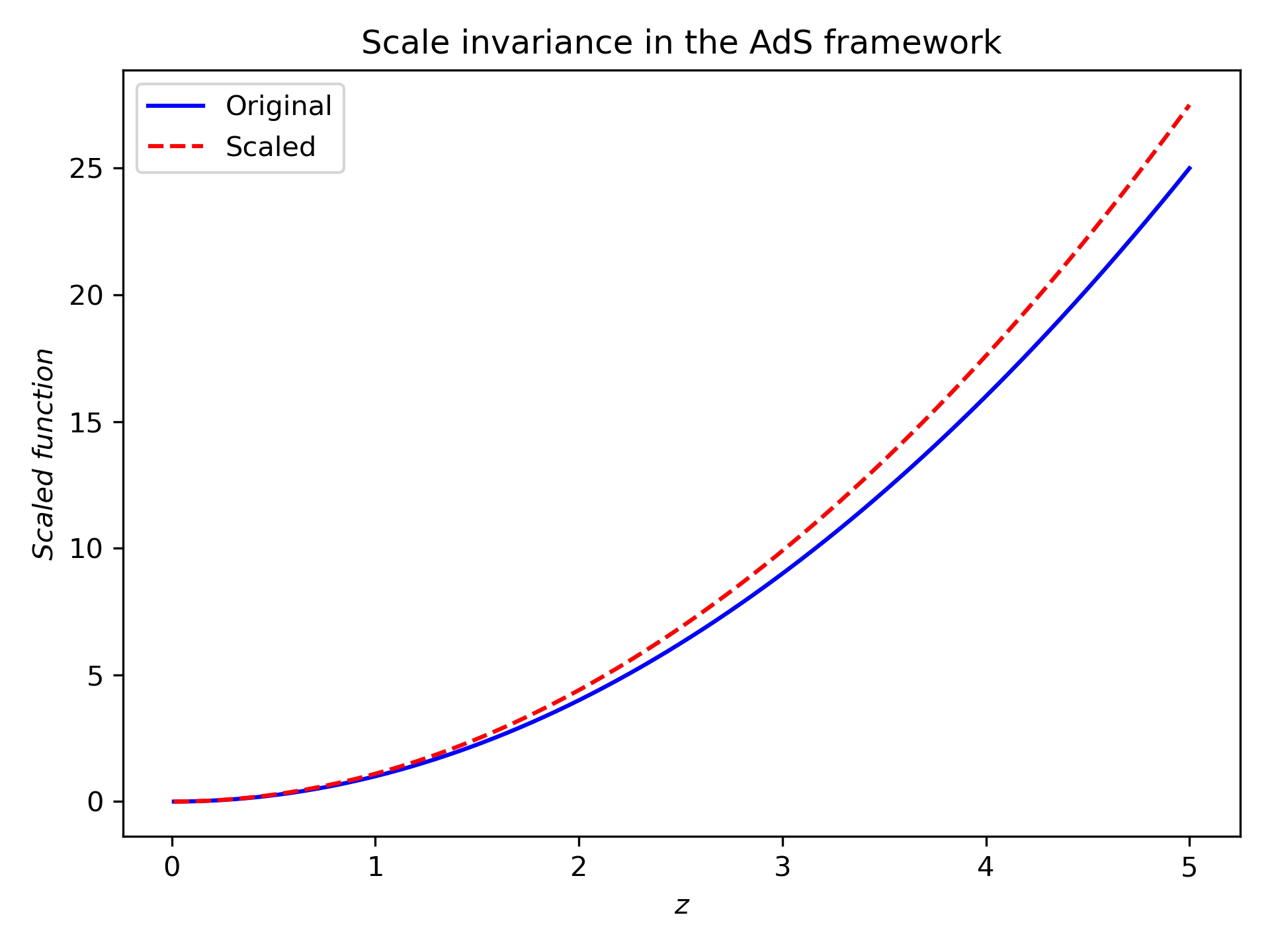}
\caption{Schematic illustration of scale invariance in the AdS framework. The figure compares the function $f(z)=z^2$ with its scaled form $f(\lambda z)/ \lambda^2$ for $ \lambda=2$, demonstrating invariance under a simple scaling transformation. The vertical axis represents a dimensionless illustrative quantity and does not correspond to a specific proton observable.}
\label{Figure3}
\end{figure}

Within this approach, scale invariance on the boundary corresponds to geometric scaling in the bulk. The invariance of the AdS metric under $x^{\mu} \rightarrow \lambda x^{\mu} , z \rightarrow \lambda z $ provides a geometric interpretation of self-similarity, precisely the kind of scaling symmetry that underlies fractal-inspired descriptions of proton structure. The light-front wavefunctions obtained from light-front holographic QCD display patterns that are self-similar under changes of the resolution scale, echoing the recursive behavior of parton distributions observed in phenomenological fractal-inspired models (see Figure \ref{Figure3}). Consequently, the holographic mapping between bulk geometry and boundary dynamics may reflect analogous scaling properties. It should be stressed, however, that this correspondence is interpretive rather than derivational. At present, there is no explicit holographic construction that yields fractal exponents, discrete scale invariance, or self-similar PDFs directly from the bulk dynamics. Instead, the comparison rests on identifying analogous scaling structures in the two descriptions. Recent developments have extended holographic models to include gluonic components, spin effects, and small \textit{x} behavior \cite{Dosch, HLFHS}, demonstrating that the rise of the proton structure function at small \textit{x} has been explored in holographic frameworks and can be qualitatively reproduced in certain model implementations. These studies suggest that certain scale-dependent features of boundary distributions can be discussed within a multi-scale geometric framework in the bulk. Moreover, later refinements of the soft-wall model and its effective potentials have further improved consistency with QCD observables, reproducing hadron mass spectra and the trace anomaly with greater accuracy \cite{Bayona}. Although the two approaches originate from different assumptions, comparing them highlights several common scaling features. Scale invariance in proton structure can be viewed, within holographic models, through the geometric scaling properties of AdS space.\\

The light-front holographic framework thus offers a coherent picture in which the proton's scale-invariant properties emerge from the geometry of the underlying dual space. By translating self-similar behavior in parton distributions into geometric scaling along the AdS coordinate, it unites phenomenological fractal-inspired models with the field-theoretic structure of QCD. This perspective sets the stage for the interpretive comparison presented in the following section.
 
 \section{Interpretive synthesis: Self-similarity and holographic scaling}

 The self-similar description of proton structure developed earlier provides a quantitative language for analyzing the scale-invariant behavior of parton distributions \cite{Rabemananjara}. Its power-law form reproduces the experimentally observed rise of the structure function $F_2(x, Q^2)$ at small \textit{x} \cite{H1}, reflecting the recursive cascade of gluons and sea quarks that dominate in this regime. At the same time, light-front holographic QCD offers a geometric realization of scale invariance in which the fifth-dimensional AdS coordinate \textit{z} encodes the inverse energy scale of the boundary theory \cite{HLFHS, Ahmady}. The comparison suggests that statistical self-similarity and geometric scaling provide complementary ways of describing scale-dependent behaviour. The discussion presented in this section is interpretive in nature. Its purpose is not to establish a formal correspondence between self-similar models and holographic QCD, but rather to examine possible conceptual similarities in how scale dependence is represented within the two frameworks.\\
 
Table \ref{Table1} summarizes this correspondence at a qualitative level. The entries in the table should not be interpreted as the outcome of a formal derivation, but rather as heuristic identifications that highlight analogies between key quantities appearing in self-similar models and their counterparts in light-front holographic QCD. Certain quantities appearing in phenomenological self-similar models can be compared, at a qualitative level, with quantities that play analogous roles in holographic descriptions.The Bjorken variable \textit{x} may be compared with the longitudinal momentum fraction associated with the boundary coordinate, while the probe virtuality $Q^2$ is often associated, in a schematic sense, with the inverse square of the holographic coordinate. The scaling exponents that govern the \textit{x} and  $Q^2$ dependence of the structure functions translate into conformal or effective scaling dimensions of the corresponding boundary operators in light-front holographic QCD. Within this interpretive framework, the rise of structure functions at small \textit{x} can be understood as a manifestation of scale invariance operating across multiple momentum scales. In self-similar models, this invariance is statistical in nature and encoded in fractal exponents determined through fits to experimental data. In contrast, holographic models express scale invariance through the geometry of the AdS metric, with confinement introduced via effective potentials that break exact conformal symmetry. The fact that both approaches emphasize scale-dependent organization suggests that they may be capturing complementary aspects of the same underlying dynamics, even though their quantitative implementations differ significantly. \\

It is also important to recognize the limitations of this comparison. Global PDF analyses based on perturbative QCD evolution, as well as lattice-QCD calculations of parton moments, provide quantitatively controlled descriptions of proton structure that go beyond the scope of both self-similar and holographic models \cite{Hou, Sato, Cocuzza, Melnitchouk}. The latter should therefore be viewed as effective or exploratory frameworks, useful for highlighting patterns and organizing principles rather than for making precision predictions. In particular, the absence of a direct holographic derivation of PDFs or fractal exponents limits the extent to which self-similarity can presently be grounded in holographic QCD. Despite these limitations, the interpretive synthesis presented here serves to clarify how phenomenological scaling laws observed in DIS may be viewed through a geometric lens. By placing self-similar parameterizations and holographic scaling side by side, one gains insight into the possible role of scale invariance as a common organizing principle. The comparison presented below is intended solely as a qualitative guide to possible similarities in how scale dependence is represented in the two frameworks. Since self-similar models and holographic QCD arise from different theoretical motivations and employ different dynamical assumptions, the entries in Table 1 should be interpreted as heuristic analogies rather than formal correspondences. Their purpose is to highlight broadly analogous roles played by certain variables and scaling parameters, not to imply a derivation of one framework from the other.

\begin{table}[H]
\caption{Schematic and qualitative comparison of scaling concepts appearing in phenomenological self-similar models and light-front holographic QCD. The entries are intended as heuristic analogies and do not represent a one-to-one theoretical mapping.}

\label{Table1}
\centering
\resizebox{\textwidth}{!}{
\begin{tabular}{|c|c|c|c|}
\hline
\textbf{Physical quantity} & \textbf{Self-similar description} & \textbf{\makecell{Light-front holographic \\ QCD analogue}} & \textbf{Typical scale/reference} \\
\hline
Scaling variable & Bjorken \textit{x} & \makecell{Scale-dependent boundary dynamics \\ associated with the holographic coordinate $z$} & HERA range $10^{-5} - 0.5 $ \cite{H1} \\
\hline
Resolution scale & Probe virtuality $Q^2$ & \makecell{Inverse-squared holographic \\ coordinate $1/z^{2}$} &  $2-50 \mathrm{GeV^2} $ \cite{H1} \\
\hline
Fractal exponent & $ \lambda(Q^2) = d \ln F_2 / d \ln (1/x)$ & \makecell{Effective scaling dimension/parameter} & $\lambda \approx 0.1-0.3$ \cite{Lastovicka, AJ3, H1} \\
\hline
Scaling symmetry & \makecell{Statistical self-similarity of \\ parton cascade} & \makecell{Geometric scale invariance of\\ AdS metric} & \makecell{Foundational works linking hard \\ scattering to gauge duality \cite{Polchinski, Brodsky}} \\

\hline
\end{tabular}
}
\end{table}

The analogies summarized in Table 1 are based on the roles played by the corresponding quantities within their respective frameworks. The Bjorken variable \textit{x} and the holographic coordinate \textit{z} are both associated with the characterization of scale-dependent structure, although in fundamentally different ways. Likewise, the probe virtuality $ Q^2$ determines the resolution scale in deep inelastic scattering, whereas the bulk coordinate in holographic descriptions is commonly related to an inverse energy scale. The comparison between scaling exponents in self-similar models and effective scaling dimensions in holographic approaches is intended only to emphasize their shared role in controlling scaling behaviour. These observations are qualitative and should not be interpreted as establishing a unique theoretical correspondence.

\section{Conclusion and Outlook}

In this review, we have distinguished between established QCD results, phenomenological self-similar descriptions of proton structure, and interpretive comparisons with holographic approaches. In this context, self-similar models provide useful parametrizations of observed scaling behaviour, while holographic QCD approaches describe these ideas from a geometric perspective. Self-similar models, inspired by fractal geometry, have proven useful in describing PDFs and structure functions, particularly in the small \textit{x} region where recursive parton cascades dominate. Their success in describing key features of DIS data highlights the relevance of scale-invariant patterns in the organization of partonic dynamics. From a complementary perspective, holographic approaches to QCD offer a geometric realization of scale transformations, in which the resolution scale of the boundary theory is mapped onto an additional spatial dimension in the bulk. Light-front holographic QCD, in particular, has proven effective in capturing essential aspects of hadron spectroscopy and form factors while maintaining approximate conformal symmetry at short distances and confinement at long distances. Within this picture, scale invariance appears naturally as a geometric property of the AdS background rather than as a statistical assumption imposed on parton distributions. The central theme of this review has been to clarify how these two approaches, phenomenological self-similarity and holographic scaling, may be viewed as complementary descriptions of multi-scale structure in the proton. The correspondences discussed here are interpretive and qualitative in nature, and no claim is made that current holographic models provide a first-principles derivation of self-similar PDFs or fractal exponents. Instead, the comparison serves to highlight common structural elements, notably the prominent role played by scale invariance across a wide range of momentum scales.\\

At present, the available evidence supports only a qualitative comparison between these frameworks. Whether a deeper connection exists remains an open question. Several limitations of the present perspective should be emphasized. Quantitative descriptions of proton structure are today dominated by global PDF analyses based on perturbative QCD evolution and by lattice-QCD determinations of parton moments. In contrast, both self-similar models and holographic approaches remain effective or exploratory frameworks. Bridging the gap between these phenomenological descriptions and more rigorous QCD-based methods remains an open challenge. In particular, a more explicit treatment of gluon and sea-quark dynamics, as well as the incorporation of running coupling effects that break conformal symmetry, will be essential for future progress.\\

Despite these open issues, the synthesis presented here suggests that scale invariance provides a useful organizing principle for understanding the proton's internal structure. By viewing fractal-inspired parameterizations and holographic geometry as complementary manifestations of scale-dependent dynamics, one gains a broader perspective on how complexity and regularity coexist in strong-interaction physics \cite{Ethier, Accardi}. Future work aimed at developing quantitative holographic descriptions of parton distributions, or at embedding self-similar scaling within more complete QCD frameworks, may help to clarify the deeper origin of the scaling patterns observed in deep inelastic scattering.

\section*{Acknowledgements}
The author thanks Late Prof. Dilip Kumar Choudhury, Gauhati University, for his relentless support and invaluable guidance, which significantly enhanced the ideas on self-similarity of the structure of proton.

\section*{Data Availability}
The data that support the findings of this study are available from the corresponding author upon reasonable request.

\section*{Conflicts of Interest}
The author declares no conflicts of interest.

\section*{Funding}
This research received no external funding.

\end{document}